# Spinon Excitations in the Quasi-1D S = ½ Chain $Cs_4CuSb_2Cl_{12}$

Thao T. Tran,[a,b] Chris A. Pocs,[c] Yubo Zhang,[d,e] Michal J. Winiarski,[a,f] Jianwei Sun,[d] Minhyea Lee,[c] and Tyrel M. McQueen*[a,g]

[a] Department of Chemistry, Department of Physics and Astronomy, Institute for Quantum Matter, Johns Hopkins University, Baltimore, MD 21218, USA
[b] Department of Chemistry, Clemson University, Clemson, SC 29634, USA
[c] Department of Physics, University of Colorado at Boulder, Boulder, CO 80309, USA
[d] Department of Physics and Engineering Physics, Tulane University, New Orleans, LA 70118, USA
[e] Department of Physics, Southern University of Science and Technology of China, Shenzhen 518055, China
[f] Faculty of Applied Physics and Mathematics, Gdansk University of Technology, Narutowicza 11/12, 80-233 Gdansk, Poland
[g] Department of Materials Science and Engineering, Johns Hopkins University, Baltimore, MD 21218, USA

The spin-½ Heisenberg antiferromagnetic chain is ideal for realizing one of the simplest gapless quantum spin-liquids (QSLs), supporting a many-body ground state whose elementary excitations are fractional fermionic excitations called spinons. Here we report the discovery of such a 1D QSL in $Cs_4CuSb_2Cl_{12}$. Compared to previously reported $S$ = ½ 1D chains, this material possesses a wider temperature range over which the QSL state is stabilized. We identify spinon excitations extending at $T$ > 0.8 K, with a large $T$-linear contribution to the specific heat, $\gamma$ = 31.5(2) mJ mol$^{-1}$ K$^{-2}$ which contribute itinerantly to thermal transport up to temperatures as high as $T$ = 35 K. At $T$ = 0.7 K, we find a second-order phase transition, suggesting a weak spin-Peierls transition that is unchanged by a $\mu_0H$ = 5 T magnetic field. $Cs_4CuSb_2Cl_{12}$ reveals new phenomenology deep in the 1D QSL regime, supporting a gapped QSL phase over a wide temperature range compared to many other experimental realizations.

## I. Introduction

Quasi-1D magnets exhibit an incredibly rich variety of physics and there is much phenomenology completely unique to 1D spin-systems. In 2D and 3D magnets, competing exchange interactions are requisite for the frustration and strong quantum fluctuations that drive novel states of matter such as quantum-spin-liquids (QSLs).[1-4] As such, there are a variety of QSLs hypothesized to exist within different lattice symmetries (2D: triangular, kagomé, honeycomb, 3D: pyrochlore, diamond, hyperkagomé) and various types of magnetic exchange (Heisenberg, Kitaev, Dzyaloshinskii–Moriya).[2-14] In contrast to these higher dimensional examples, quasi-1D quantum magnets are an excellent platform for realizing QSLs without frustration. In particular, the quantum 1D S = ½ Heisenberg antiferromagnetic chain (HAFC) is an important theoretical paradigm with emergent collective behavior that realizes a gapless QSL. The idealized model, which considers only nearest-neighbor interactions:

$$H = J \sum_{\langle ij \rangle} S_i \cdot S_j, \qquad (1)$$

has an exactly solvable ground state which is a macroscopically entangled QSL state given by the Bethe Anzatz.[15-24] In this many-body state, the elementary excitation of the HAFC is a type of fractional fermionic quasi-particle called a spinon, which carries spin-½. Thermodynamically, spinons may be identified in an insulating magnet by a distinct $T$-linear contribution to the specific heat at temperatures $T \ll J/k_B$ and their field-dependent itinerant contribution to thermal conductivity.[25]

One of the first experimental realizations of the $S$ = ½ HAFC was discovered in $CuGeO_3$.[18] This material undergoes a magnetic phase transition to a long-range ordered spin-Peierls state at $T$ = 14 K.[18, 26] Another seminal example of a $S$ = ½ HAFC is $Sr_2CuO_3$.[27-29] This system exhibits very strong intra-chain exchange coupling $J/k_B$ yet also undergoes a 3D magnetic phase transition at $T \approx$ 5 K.[27] Indeed in most real quasi-1D systems, the viable temperature range over which QSL physics or spinon-like excitations may be observed is limited by the fact that inter-chain couplings will eventually stabilize 3D long-range order, or by the fact that dimerization at low enough temperatures will favor a state of localized spin-singlets supporting gapped bosonic excitations.

In this paper, we report the discovery of a QSL $S$ = ½ HAFC in the new material $Cs_4CuSb_2Cl_{12}$ (CCSC). It features a wide temperature window over which the QSL state is stabilized. We observe spinon excitations at $T$ > 0.8 K, characterized by a large $T$-linear contribution $\gamma$ = 31.5(2) mJ mol$^{-1}$ K$^{-2}$ contribution to the specific heat that extends to at least 5 K, deducing an intra-chain exchange coupling $J/k_B$ = 176(2) K. We additionally report the first evidence of a second-order phase transition at $T_c$ = 0.7 K, finding an anomaly in the specific heat that is insensitive to applied fields up to $\mu_0H$ = 5 T, suggesting the existence of a weak spin-Peierls transition into a low-$T$ partially spin-dimerized state. Further investigation of thermal transport reveals field-dependence up to as high as $T$ = 35 K associated with an itinerant magnetic contribution, suggesting that spinon-like excitations in CCSC persist up to temperatures as high as 0.2 $J/k_B$.

## II. Experiment

Polycrystalline CCSC was synthesized through solid-state reactions at 220 °C for 3 days by combining stoichiometric amounts of CsCl, CuCl$_2$ and SbCl$_3$. Dark purple triangular-shaped crystals of CCSC were grown by hydrothermal techniques. The reaction mixture of 1 g of polycrystalline CCSC and 10 mL of 12 M HCl were placed in a 23-mL Teflon-lined stainless steel autoclave. The autoclave was closed, gradually heated up to 150 °C, held for 3 days, and then slowly cooled to room temperature at a rate of 6 °C h$^{-1}$.

Powder X-ray diffraction (PXRD) data were collected at room temperature using Bruker D8 Focus diffractometer with a LynxEye detector using Cu K$\alpha$ radiation ($\lambda$ = 1.5424 Å). Rietveld refinements on PXRD data were performed using TOPAS 4.2.

Single crystal X-ray diffraction (SXRD) data were collected at $T$ = 213 K using the program CrysAlisPro (Version 1.171.36.32 Agilent Technologies, 2013) on a SuperNova diffractometer equipped with Atlas detector using graphite-monochromated Mo K$\alpha$ ($\lambda$ = 0.71073 Å). CrysAlisPro was used to refine the unit cell dimensions and for data reduction. The temperature of the sample was controlled using the internal Oxford Instrument Cryojet. The structure was solved using SHELXS-97 and refined using SHELXL-97. All calculations were performed using SHELXL-97 crystallographic software package.[30]

Powder neutron diffraction data were collected using the time-of-flight high-flux NOMAD at the Spallation Neutron Source, Oak Ridge National Laboratory. Rietveld refinements were performed using GSAS in EXPGUI.[31]

Physical property characterization was performed using Quantum Design Physical Properties Measurement System (PPMS). Magnetization data were collected using the VSM options at $T$ = 2 – 300 K under $\mu_0H$ = 5 T and converted to magnetic susceptibility using the approximation $\chi = M / H$. Heat capacity data were collected using the semiadiabatic pulse technique (2% heat rise) for $T$ = 0.2 – 300 K.

Density functional calculations are carried out by using the projector augmented-wave method implemented in the Vienna ab initio simulation package (VASP).[32-33] The recently developed strongly constrained and appropriately normed (SCAN) density functional is used to treat the exchange-correlation interactions.[34] SCAN functional is able to stabilize the magnetic moment on Cu, while the conventional local density approximation (LDA) and generalized gradient approximation (GGA) fail to do so. An energy cutoff of 500 eV is to truncate the plane-wave basis, and a 4 × 4 × 4 K-mesh is used in sampling the Brillouin zone of 76-atom cell. All calculations are done with the experimental crystal structure.

## III. Magnetism:

The structure of CCSC consists of $S$ = ½ CuCl$_4$ plaquettes connected by Sb ions to form 1D chains, Fig. 1. The magnetic susceptibility, measured at $\mu_0H$ = 5 T and estimated as $\chi \approx M/H$, shows a broad maximum around $T$ = 210 K, followed by an upturn below $T$ = 50 K. This behavior is characteristic of a $S$ = ½ HAFC in the presence of a few magnetic impurity spins.[15, 18, 21, 27, 35] No indication of a transition to a 3D long-range-ordered magnetic state is observed in magnetization measurements at temperatures $T$ > 2 K. Quantitative analysis was performed by using the $S$ = ½ HAFC Bonner-Fisher model plus a Curie-Weiss term for the defect spins:[17, 36-38]

$$\chi = \chi_{BF}(J,T) + C/(T-\theta), \qquad (2)$$

where $\chi$ is the observed susceptibility, $J$ is the intrachain exchange interaction, $C$ and $\theta$ are the Curie constant and Weiss temperature for the defect spins. An optimized fit to the data gives $J/k_B$ = 186(2) K, $C$ = 0.011(1) emu mol-f.u.$^{-1}$ Oe$^{-1}$ and $\theta$ = -37(5) K. The chain interaction strength is consistent with location of the observed maxima of $\chi$ and additionally consistent with an alternative fit utilizing a dimer model (not shown). The Curie constant corresponds to ~3% of $S$ = ½ impurity spins, and the corresponding Weiss temperature indicates a larger interaction strength than would be expected between isolated $S$ = ½ units in free space, consistent with these impurity spins being embedded in a QSL.

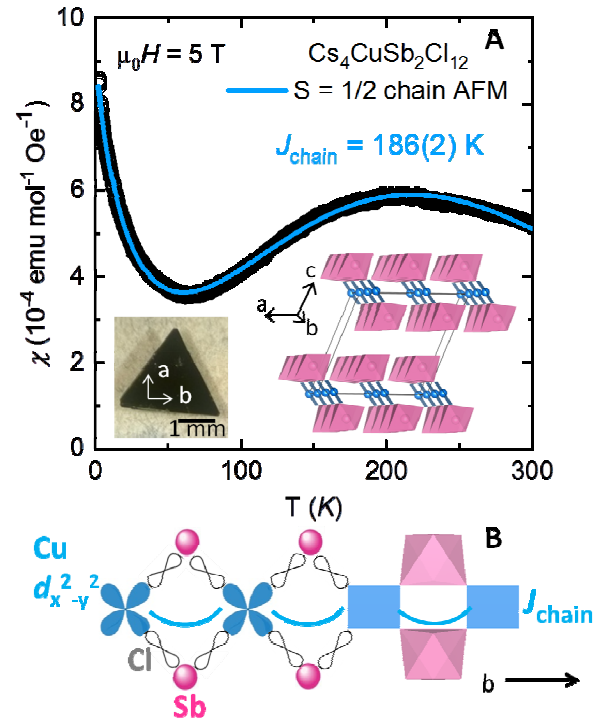

**Figure 1.** (A) The magnetic susceptibility of Cs$_4$CuSb$_2$Cl$_{12}$ as a function of temperature. A S = ½ HAFC model gives an intrachain exchange coupling of $J/k_B$ = 186(2) K; Inset: single crystal and overall crystal structure; (B) The structure of the 1D chains of Cs$_4$CuSb$_2$Cl$_{12}$, showing the super-exchange interactions along the b-axis.

The broad maximum feature of the magnetic susceptibility of Cs$_4$CuSb$_2$Cl$_{12}$ at $T$ ~ 210 K (Fig. 1A) could indicate that the spins were already paired up at high temperature $T > 300$ K. In order to unambiguously rule out this possibility, time-of-flight neutron powder diffraction (TOF-NPD) measurements were performed at $T = 300$ $K$ by NOMAD. The TOF-NPD patterns were simultaneously analyzed by Rietveld refinements in order to determine the presence or absence of magnetic ordering. The data were fit well with the crystal structure of $Cs_4CuSb_2Cl_{12}$ and no additional Bragg peaks attributed to magnetic ordering of any kind are observed. As a test, a magnetic phase was added to the refinement to assess whether magnetic scattering would be visible. To estimate the sensitivity to magnetic order, we used AFM state. This yielded an upper limit on the magnetic moment of 0.2(5) $\mu_B$ per Cu.

To further explore the exchange coupling in this system and verify its quasi-1D nature as a $S = ½$ HAFC, we performed density functional theory calculations using the recently developed strongly-constrained and appropriately-normed (SCAN) exchange correlation functional.[39] We used CCSC crystal structure determined from TOF-NPD and X-ray diffraction measurements, Table 1.[40]

**Table 1**: Crystallographic data for $Cs_4CuSb_2Cl_{12}$, S.G. $C2/m$ (No. 12), $a = 13.083(3)$Å, $b = 7.3507(2)$Å, $c = 13.070(3)$Å, $\beta = 112.17(3)°$.

|     | X          | y         | z          | s.o.f | Uiso      |
| --- | ---------- | --------- | ---------- | ----- | --------- |
| Cs1 | 0.1241(2)  | 0.0000    | 0.3698(2)  | 1     | 0.0229(9) |
| Cs2 | 0.3745(2)  | 0.0000    | 0.1199(2)  | 1     | 0.0243(9) |
| Sb1 | 0.2520(2)  | 0.5000    | 0.2586(2)  | 1     | 0.0127(9) |
| Cu1 | 0.0000     | 0.0000    | 0.0000     | 1     | 0.0142(2) |
| Cl1 | 0.1119(5)  | 0.2206(8) | 0.1106(5)  | 1     | 0.0243(2) |
| Cl2 | 0.3739(6)  | 0.5000    | 0.1353(6)  | 1     | 0.0238(2) |
| Cl3 | 0.3693(4)  | 0.2626(7) | 0.3728(4)  | 1     | 0.0239(2) |
| Cl4 | 0.1345(6)  | 0.5000    | 0.3769(6)  | 1     | 0.0239(2) |

Results of our calculations are shown in Fig. 2. The magnetic density on Cu sites clearly reveals that $d_{x^2-y^2}$ orbitals are polarized, direct evidence of $3d^9$ $Cu^{2+}$ cations with effective $S = ½$. Within the $ab$-plane, the spin of $Cu^{2+}$ also polarizes the $p_x$ and $p_y$ electrons on the four Cl ligand anions, forming $CuCl_4$ units with AF couplings along the $b$-axis. We have observed similar spin polarization of oxygen in cuprates.[34, 41] To estimate the strength of the exchange coupling $J$, the total energies of the antiferromagnetic (AFM) and ferromagnetic (FM) states of the nearest-neighbor $S = ½$ Heisenberg Hamiltonian in the mean-field approximation were calculated. The difference in total energies of the AFM and FM phases is given by:

$$\Delta E = E_{AFM} - E_{FM} = JNZ\langle S \rangle^2, \quad (4)$$

where $N$ is the total number of magnetic moments, $Z$ is the number of nearest-neighbor spins, and $\langle S \rangle$ is the averaged spin on each site. Because the spin density supports the predominance of intra-chain interactions, we take $Z = 2$. These calculations were normalized to one formula unit, so $N = 1$. Our first-principles computations for the total energies of the AFM and FM states resulted in a predicted $J/k_B \sim 200$ K (depending on the precise approximations used), lending further plausibility to the experimentally determined $J/k_B$ value of 186(2) K deduced from susceptibility fits, and our analysis of the specific heat.

We calculate a Cu magnetic moment of 0.52 $\mu_B$. Note that there is also considerable contribution to the magnetization from Cl ligands, with each Cl contributing an additional magnetic moment of 0.06 $\mu_B$. The theoretical bandgap calculated is 0.28 eV, smaller than the experimental value of ~ 1 eV estimated from UV-Vis spectroscopy measurements. The valence band maximum is mostly composed of the Cu-$d_{x^2-y^2}$, Cl-$p$ and Sb-$s$ states. The conduction band minimum is isolated from other bands, and mainly derived from Cu $d_{x^2-y^2}$ states with some hybridized Cl-$p$ states. Thus DFT calculations also support CCSC as a realization of a 1D HAFC.

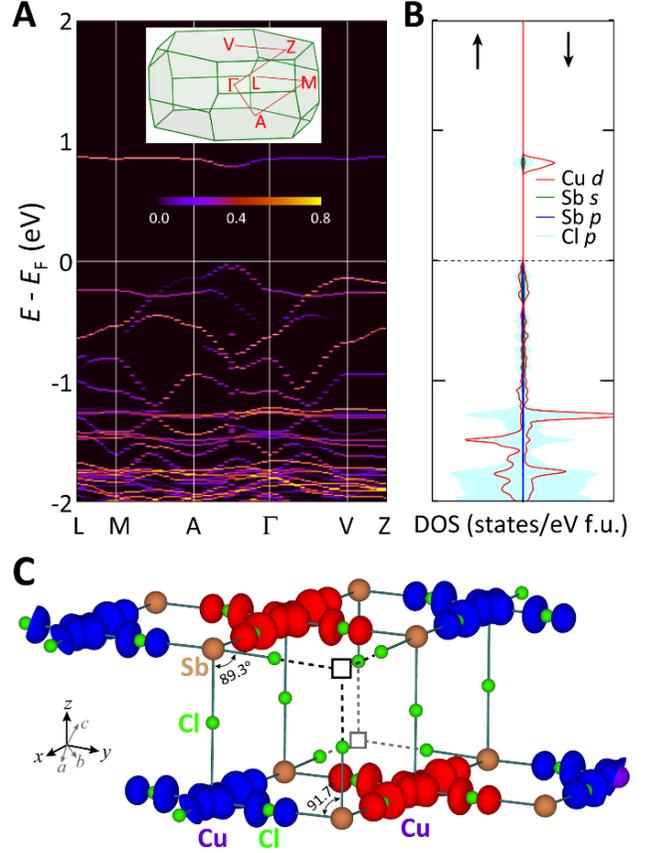

**Figure 2.** (A) Unfolded band structure of $Cs_4CuSb_2Cl_{12}$ AFM 1-D chain. The unfolding maps the band structure of the 2×2×1 AFM supercell into that of the 1×1×1 non-magnetic primitive cell, and a weaker spectral weight (purple color on the left end of the color bar) represents the band structure is more significantly affected by the magnetism. The inset shows the band path in the primitive cell Brillouin zone. (B) Spin-polarized density of states (DOS). (C) Calculated magnetic density. The red and blue isosurfaces denote the spin up and spin down charge densities, respectively. Cu ions form AFM chains within the $xy$-plane, but they do not have magnetic coupling across the planes. The open squares (☐) denote empty sites.

### IV. Heat Capacity:

To elucidate the thermodynamics of the ground state, we performed heat capacity measurements over the range 0.2 K $\leq T \leq$ 300 K.

For $T \geq 2$ K, there are no sharp anomalies, indicative of a lack of long-range magnetic ordering or other phase transitions (Fig. 3). The heat capacity data above T = 2 K are modeled very well with one Einstein mode ($E(\theta_E, T)$), one Debye mode ($D(\theta_D, T)$) and electronic contribution ($\gamma$) following the equation:

$$C_p/T = E(\theta_E, T)/T + D(\theta_D, T)/T + \gamma \quad (5)$$

The Einstein and Debye temperature can be extracted from the fit to be 61(2) K and 178(7) K, respectively.

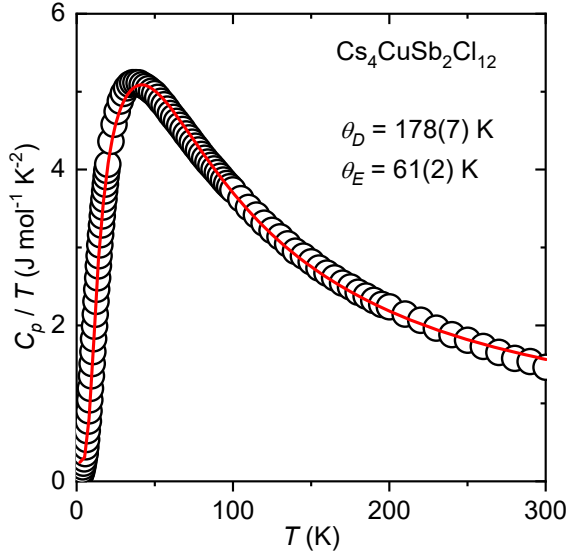

**Figure 3.** The specific heat $C_p/T$ vs $T$ of CCSC from $T$ = 2 K to 300 K shows no sharp anomalies, indicative of a lack of long-range magnetic ordering or other phase transitions

There is, however, an apparent $T$-linear contribution that extends up to at least $T$ = 5 K. To further explore this contribution, expected for a 1D HAFC QSL,[17, 42] we performed specific heat measurements in a Quantum Design PPMS dilution refrigerator from $T$ = 0.2 K to $T$ = 3 K, Fig. 4.

$T$-power-law behavior of the heat capacity at low temperatures is deeply informative about the nature of low-energy quasi-particle excitations. The power law exponent is not only telling of the dimensionality of the excitations, which modifies the density of states, but also indicative of their dispersion $\varepsilon(k)$ at zone-center, as for bosonic and fermionic statistics alike, contributions from all other parts of the excitation spectrum are exponentially suppressed as $\exp(-\varepsilon/k_B T)$ at low $T$. In virtually every insulator, the most obvious low-$T$ contribution to the heat capacity is a $T^3$ phononic term, describable via the Debye model. In a magnetic insulator, the presence of a large $T$-linear contribution is strongly indicative of low-dimensional magnetic excitations, and only a few combinations of dimensionality and low-$k$ dispersion can theoretically produce precisely $C \sim T$. In the context of CCSC, 1D spinons are the only plausible source of such a sizable linear contribution. The predicted $T$-linear spinon contribution intrinsic to the HAFC QSL is given by:[43,45]

$$\gamma = \frac{2}{3} R k_B / J \quad (6)$$

where $J$ is the AFM intrachain exchange interaction strength. Between 0.7 K < T < 3 K, we find that the total specific heat is well-modeled by the following the equation, as shown in Fig 4. (a):

$$C_p/T = \gamma + \beta_3 T^2, \quad (7)$$

including terms for the expected magnetic ($\gamma$) and phononic ($\beta_3$) contributions to the specific heat. An optimized fit yields values of $\gamma$ = 31.5(2) mJ mol$^{-1}$ K$^{-2}$, and $\beta_3$= 7.74(5) mJ mol$^{-1}$ K$^{-4}$, respectively. From the above formula for $\gamma$, we deduce an exchange coupling $J/k_B$ = 176(2) K, which is in excellent agreement with the $J/k_B$ = 186(2) K determined from the magnetic susceptibility.

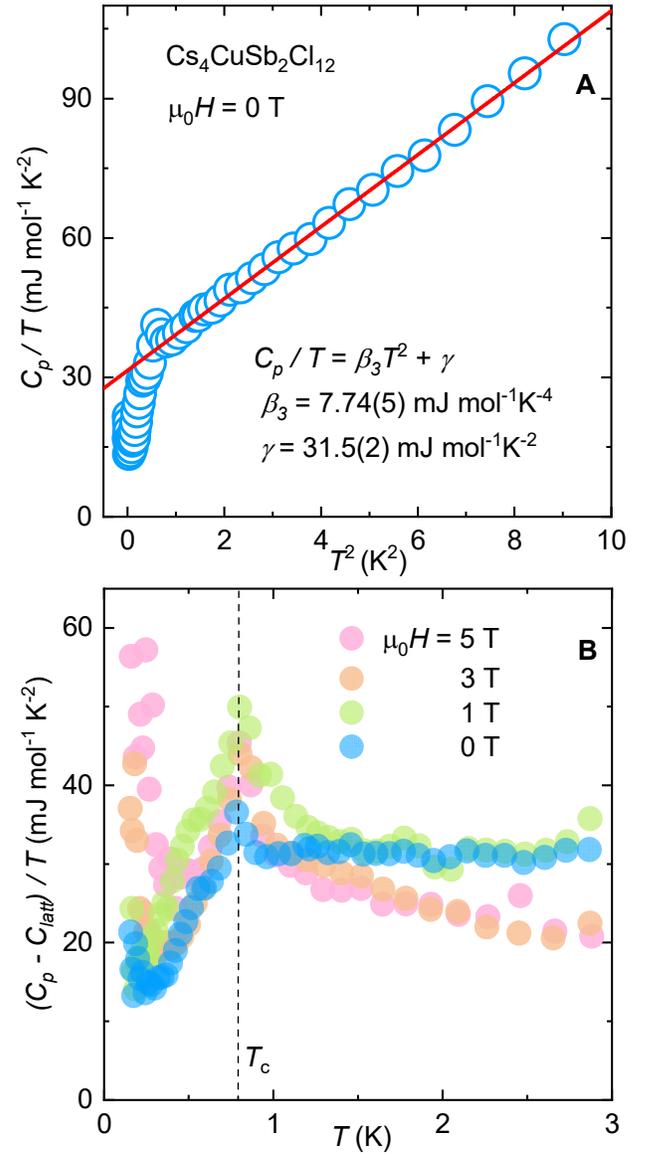

**Figure 4.** (A) Low temperature heat capacity of CCSC $C_p T^{-1}$ vs. $T^2$ plot shows the existence of a linear correlation from which phononic and magnetic contribution can be extracted; (B) ($C_p - C_{latt}$) / $T$ vs. $T$ plot depicts the second-order phase transition.

This $T$-linear term is disrupted by a phase transition at $T_c = 0.7$ K. The $\lambda$-like anomaly is characteristic of a second-order phase transition. The concomitant loss of the majority of the $T$-linear contribution suggests that spinon excitations are partially or entirely gapped occurring as spins of the 1D chain dimerize into a spin-Peierls state. The dimerized state is characterized by the organization of the spins into nearest-neighbor singlet pairs, the fundamental excitations of which are gapped bosonic triplet states. Application of a $\mu_0 H = 5$ T magnetic field does not suppress or move the transition within the resolution of our measurements, suggesting that the spin-Peierls transition is weakly coupled to the applied field. Taken together, these results imply the $\gamma$ term is intrinsic to CCSC, not due to disorder or impurity spins, and necessarily a consequence of fermionic spinon excitations, i.e., gapless excitations in the QSL state.[16, 23, 44-45]

## V. Thermal Transport:

The in-plane longitudinal thermal conductivity $\kappa_{xx} \equiv \kappa$ was measured on an as-grown sample of dimension approximately $1 \times 0.5 \times 0.2$ mm$^3$, using a single-heater, two-thermometer configuration in steady-state operation with magnetic field applied either in the ab-plane ($H||\nabla T||ab$) or normal to the plane ($H\perp ab$). Over the entire temperature range of measurement, the difference in absolute temperatures across the sample was set so that it never exceeded 5% of the bath temperature. All thermometry was performed using CERNOX chip resistors, which were pre-calibrated individually and in-situ, both with and without applied fields.

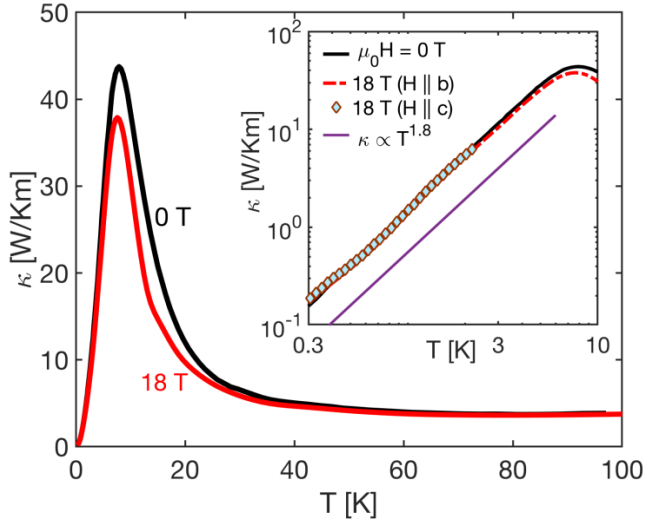

**Figure 5**. $\kappa$ shown as a function $T$ for 0 T (ZF) and 18 T applied normal to the plane. Temperature gradient was in the $ab$-plane. Inset $\kappa$ ($T$) shown on a logarithmic scale for comparison of the low-$T$ data at different fields. No sign of transition at $T_c \sim 0.7$ K has been observed.

Fig.5 shows $\kappa$ as a function of temperature for zero field (ZF) and the maximum field applied $\mu_0 H = 18$ T. We checked both in-plane and out-of-plane orientations of field and found that the data at finite field is completely insensitive to the direction of the applied field. For ZF and 18 T, conspicuous peaks of similar size occur at the same temperature, approximately 8 K, followed by rapid decrease of $\kappa$, as $T$ is further lowered even below $T_c \sim 0.7$ K. At the peak, the suppression in $\kappa$ is about 10 % of the ZF value and it is observed only in 3 K$\lesssim T \lesssim$ 35 K. At low temperatures, $T<$ 3 K, $\kappa$ becomes robustly field-independent up to 18T, and follows a power law-dependence with approximate exponent $\kappa \sim T^{1.8}$. Our measurement bears many similarities to thermal conductivity data measured in the S = ½ 1D HAFC chain cuprate Sr$_2$CuO$_3$, namely (1) thermal conductivity in the low-$T$ ballistic regime also follows an apparently sub-quadratic $T$-power law, and (2) despite a significant $T$-linear contribution to the heat capacity in the disordered QSL phase, an itinerant spinon contribution does not apparently have a significant effect on bulk thermal conductivity until $T$ is much larger than the ordering temperature that disturbs that a QSL phase.[28, 43] These observations can be contrasted with the naive formulation that the measured thermal conductivity may be separated into distinct phonon and itinerant spinon contributions:

$$\kappa = \kappa_{ph} + \kappa_{sp}, \quad (8)$$

from which we might expect a sum of distinct $T^3$ and $T$-linear contributions $\frac{1}{3}C_{latt}v_{ph}l_{ph}$ and $C_m v_{sp} l_{sp}$ respectively in the ballistic regime. This formulation implies a short spinon-phonon relaxation time, $\tau_{sp-ph}$, compared to the intrinsic phonon relaxation time $\tau_{ph} = l_{ph}/v_{ph}$, in order for spin excitations to effectively thermalize with the underlying lattice. In fact only $\kappa_{ph}$ becomes experimentally measurable in the limit $\tau_{sp-ph} \ll \tau_{ph}$. Paradoxically however, the clear separation of $\kappa$ into fully independent lattice and magnetic terms implies that they contribute distinctly, when in reality they are necessarily intertwined by the spin-phonon interaction, quantified by $\tau_{sp-ph}^{-1}$, that renders a magnetic contribution measurable in the first place. In this regard, the spin-phonon coupling may be thought of as completely necessary to probe itinerant magnetic heat transport, yet simultaneously responsible for deviations from simple theory and hybridization of phonon-spinon modes. Given a finite spin-phonon relaxation time, eqn. (8) may be modified as follows:

$$\kappa = \tfrac{1}{3} C_{latt} v_{ph} [l_{ph,0}^{-1} + (v_{ph}\tau_{sp-ph})^{-1}]^{-1}$$
$$+ C_m v_{sp} [l_{sp,0}^{-1} + (v_{sp}\tau_{sp-ph})^{-1}]^{-1}, \quad (9)$$

where $l_{ph,0}$ and $l_{sp,0}$ are mean free paths intrinsic to phonons/spinons respectively. $v_{ph}$ and $v_{sp}$ are the average speed of sound and spinon-velocity respectively, which are both estimated to be of the order $10^3 - 10^4$ m/s, given the similar size of the Debye temperature and exchange constant of CCSC. Field-independence in the lower-$T$ region and lack of indication of the magnetic transition indicates that the thermal transport is likely either phonon-dominated, or that we are insensitive to itinerant magnetic degrees of freedom in this region. In eqn. (9) such a scenario may arise provided that $l_{sp,0} \ll l_{ph,0}$, in which case the second term may be neglected. However, the unusual exponent in this ballistic regime $T^{1.8}$ (inset of Fig. 4), suggests an indirect role of spin excitations in influencing the bulk heat transport. While the boundary limited phonon mean-free path, $l_{ph,0}$, is likely to be temperature independent in the low-$T$ regime, strong temperature dependence $l_{ph}(T)$ can arise if the spin-phonon interaction is sensitive to the absolute density of magnetic excitations, i.e., the thermal population of spinons that may scatter phonons. Even in the case that thermally excited spinons are only coherent over small length scales ($l_{sp,0} \ll l_{ph,0}$), and therefore not contributing itinerantly to thermal transport, they may still modify the lattice thermal conductivity or hybridize with phonon modes provided that $v_{ph}\tau_{sp-ph}$ and $l_{ph,0}$ are comparable. If the time/length-scale for spin-phonon interaction is much smaller than that for boundary limited collisions, $v_{ph}\tau_{sp-ph} \ll l_{ph,0}$, i.e., spin-phonon interaction sets the shortest scale for phonon scattering, the phonon thermal conductivity can be further approximated as:

$$\kappa_{ph} = \tfrac{1}{3} C_{latt} v_{ph}^2 \tau_{sp-ph}. \qquad (10)$$

In this limit, which is valid provided that $l_{ph,0}$ is approximately ~ 1 mm (the length of the sample in the direction of $\nabla T$) the spin-phonon relaxation at $T = 2$ K is estimated to be $\tau_{sp-ph} \sim 40$ ns, or equivalently a phonon scattering length of ~ 100 μm. Given the experimental regime $l_{sp,0} \ll v_{ph}\tau_{sp-ph} \ll l_{ph,0}$ suggested by the data, temperature dependence of $\tau_{sp-ph}(T)$ is likely the main factor responsible for the deviations of $\kappa(T)$ from the $T^3$ behavior of $C_{latt}$.

At intermediate temperatures, 3 K < $T$ < 35 K, it is plausible that 1D HAFC spinons with a gap-less linear dispersion, contribute to coherent heat transport. The suppression near the peak with increasing applied field may be attributed to the coexistence of both types of heat carrier, phonons and spinons, over similar length scales. An applied field can gap spin excitations, leading to the reduction of spinon populations and overall $\kappa$. The spinon contribution finally becomes insignificant at $T$ > 35 K, where the field dependence disappears once again. This finding suggests that spinon-like excitations must persist up to at least $T \sim 0.2 \, J/k_B$.

## VI. Conclusions:

CCSC harbors a nearly ideal HAFC QSL state, with a weak spin-Peierls transition at $T = 0.7$ K. CCSC has an exceptionally large temperature window over which the QSL state is stable, as judged by the ratio of the intrachain interaction strength to the ordering or phase transition temperature, Table 2. KCuGaF$_6$ was also realized as a $S = ½$ HAFC with $J/k_B = 100$ K.[43] However, the nuclear Schottky contribution becomes overwhelming at low temperature, hindering study of the physics deep in the QSL phase.[43] LiCuSbO$_4$ features complex frustrated 1D chain, which cannot be solely described by the HAFC model.[15] Bi$_6$V$_3$O$_{16}$ was reported to exhibit a $S = ½$ uniform HAFC with $J/k_B = 100$ K, nevertheless, further knowledge of a possible ordering temperature deep in the 1D QSL regime is needed.[46] Of many extensively studied S = ½ HAFC systems, CCSC exhibits the largest factor of $(J/k_B) / T_c$, indicative of the wide temperature window over which the QSL is stabilized. This is further supported by the apparent effect of spinons on thermal transport, contributing itinerantly at temperatures as high as $T = 35$ K, and potentially setting the shortest time-scale for low-$T$ phonon-scattering.

**Table 2:** A summary of the exchange interaction $J/k_B$, transition temperature $T_c$ and normalized factor $(J/k_B) / T_c$ of Cs$_4$CuSb$_2$Cl$_{12}$ and the other $S = ½$ HAFCs.

| Compound | $J/k_B$(K) | $T_c$(K) | $(J/k_B)/T_c$ | Ref. |
|---|---|---|---|---|
| Cs$_4$CuSb$_2$Cl$_{12}$ | 186 | 0.7 | 270 | This work |
| KCuMoO$_4$(OH) | 238 | 1.5 | 160 | 21 |
| Sr$_2$CuO$_3$ | 2200 | 20 | 110 | 27 |
| Nd$_2$CuO$_4$ | 156 | 1.5 | 104 | 35 |
| BaCu$_2$Ge$_2$O$_7$ | 540 | 8.8 | 61 | 47 |
| Sm$_2$CuO$_4$ | 189 | 5.9 | 32 | 35 |
| BaCu$_2$Si$_2$O$_7$ | 280 | 9.2 | 30 | 20 |
| VOSb$_2$O$_4$ | 245 | 14 | 18 | 48 |
| TiOCl | 660 | 67 | 10 | 49 |
| CuGeO$_3$ | 88 | 14 | 6.3 | 18 |
| CuSb$_2$O$_6$ | 48.2 | 8.5 | 5.7 | 50 |
| CuCl$_2$ | 90 | 24 | 3.8 | 51 |
| Cs$_2$CuCl$_4$ | 2.0 | 1 | 2.0 | 52 |

In summary, we have provided the first evidence of spinon QSL physics in the new insulating material CCSC. This realization of a QSL in this $S = ½$ HAFC was deduced from the results of magnetic susceptibility, specific heat, neutron diffraction, thermal transport, and DFT computations. CCSC differs from other quantum HAFC systems for the following reasons, (i) the temperature window over which a QSL is stabilized is large compared to other examples, (ii) spinon excitations extending from $T > 0.8$ K with a large $\gamma = 31.5(2)$ mJ mol$^{-1}$ K$^{-2}$ contribution are identified and they continue to play a role in thermal transport up to $T \sim 35$ K, and (iii) a second-order weak spin-Peierls phase transition is observed that is insensitive to applied fields up to $\mu_0 H = 5$ T, suggesting a robust spin-stiffness in the vicinity of the phase transition. Additional characterizations including ultrasound and inelastic neutron scattering are underway to help gain a better understanding of the underlying physics and its unique ability to stabilize QSL over such wide temperature range in this system.

**Acknowledgements**
This work was supported as part of the Institute for Quantum Matter, an Energy Frontier Research Center funded by the U.S. Department of Energy, Office of Science, Office of Basic Energy Sciences, under Award No. DE-SC0019331. We acknowledge the support from Johns Hopkins University, Krieger School of Art and Sciences, Faculty Innovation Award, and the Johns Hopkins University Discovery Award. The dilution refrigerator was funded by the National Science Foundation, Division of Materials Research, Award #0821005. The research at Oak Ridge

National Laboratory's Spallation Neutron Source was sponsored by the U.S. Department of Energy, Office of Basic Energy Sciences, Scientific User Facilities Division. J.S and Y.Z. were supported by the US Department of Energy under EPSCoR Grant No. DE-SC0012432 with additional support from the Louisiana Board of Regents. M.J.W was supported by the Foundation for Polish Science (FNP). T.T.T acknowledges the support from Clemson University (start-up funding). C.A.P. acknowledges the support of the Colorado Energy Research Collaboratory. A portion of this work was performed at the National High Magnetic Field Laboratory, which is supported by National Science Foundation Cooperative Agreement No. DMR-1644779 and by the State of Florida.

**Keywords:** S = ½ 1D chain • Phase transition • Quantum spin liquid